\documentstyle[prd,aps]{revtex}
\def\RE {I\kern-6pt R    }
\def\Z  {Z\kern-13pt Z   }

\def\bi {\begin{itemize} }

\def\ei {\end{itemize}   }
\def\gtwid{\mathrel{\raise.3ex\hbox{$>$\kern-.75em\lower1ex\hbox{$\sim$}}}}
\def\ltwid{\mathrel{\raise.3ex\hbox{$<$\kern-.75em\lower1ex\hbox{$\sim$}}}}
\input epsf
\begin{document}

\title{A five-dimensional Schwarzschild-like solution}

\author{R. Steven Millward\footnote {Electronic address:  
                                     \tt rsmillward@byu.edu\hfil}
       }
\address{Department of Physics and Astronomy,
         Brigham Young University,
         Provo, Utah 84604}

\maketitle

\begin{abstract}
In recent years, interest in extra dimensions has experienced a dramatic increase.  A common practice has been to 
look for higher-dimensional generalizations of four-dimensional solutions to the Einstein equations.  In this vein,
we have found a static, spherically symmetric solution to the 5-d Einstein equations.  Certain aspects of this 
solution are very different from the 4-d Schwarzschild solution.  However, in observationally accessible regions,
the geodesics of the two solutions are essentially the same.      
\end{abstract}

\section{Introduction}
\label{sec:introduction}
In 1747, Immanuel Kant published ``Thoughts on the True Estimation of Living Forces", possibly the first exposition
on the existence of extra dimensions \cite{Halpern}.  His argument in favor of higher dimensional worlds was largely 
theological.  Nevertheless, it helped set the stage for the scientific consideration of such ideas.  The existence of
an extra dimension was first realized through the amalgamation of space and time into the four-dimensional entity, spacetime.
Indeed, one of the key revelations of special relativity is that our universe is four-dimensional.  

Less than a quarter of a century after the creation of special relativity Kaluza and Klein \cite{KaKl} postulated the
existence of an extra spatial dimension.  Their motivation for doing so was to give a unified description of 
electromagnetism and gravity in terms of a five-dimensional metric.  While this was only partially successful (the 
existence of the dilaton has not been verified) it gave rise to a more contemporary unification scheme; string theory.
String theory posits the existence of six or seven extra spatial dimensions.  In order not to conflict with current
experimental evidence, these extra dimensions are typically considered to be compactified into some Calabi-Yau
structure, the size of which is small enough to escape current observation.

In 1996, Horava and Witten discovered an eleven-dimensional solution to the strong coupling limit of $E_8 \times E_8$
heterotic string theory \cite{HorWit}.  In this theory all of the Standard Model fields are constrained to exist on
a ten-dimensional hypersurface dubbed the ``brane."  Gravitons, on the other hand, can migrate off of the brane and
into the ``bulk."  Thus gravity can ``leak" into the hidden dimension.  In fact, this gravitational leaking has been
used to give a phenomenological explanation of the hierarchy problem in particle physics \cite{ADD}.  Offshoots of
this theory, in which our four-dimensional universe is simply a brane in a higher dimensional bulk, are referred to
as ``braneworld" models. Such models include those of Randall and Sundrum \cite{RanSun} in which the bulk is warped
as well as the work of Dvali {\it et. al.} \cite{DvGaPo} in which the bulk dimension is infinite.

What is similar about the different braneworld models mentioned above is the character of the extra dimension.  All
of the models assume that the bulk dimension is spacelike.  While this assumption has certainly led to many 
interesting discoveries, the question remains as to the differences encountered if the bulk dimension is assumed to be
timelike.  Indeed, as was pointed out in \cite{KoWi} and the references therein, the introduction of a second time is not
without consequence.  This introduction leads to (1) a wrong sign for the Maxwell action in Kaluza-Klein type theories, (2)
the existence of tachyons, and (3) the existence of closed timelike curves.  However, if 
the extra time is considered as a bulk dimension, inaccessible to material particles, (2) and (3) no longer hold.  
Further, if the intent is to merely investigate the presence of an additional time with no reference to unification,
(1) is no longer applicable.

The inclusion of a second timelike variable is not new.  It has been investigated extensively by Bars \cite{Bars} in
the context of finding hidden symmetries in dynamical systems with one time.  Horwitz, along with Piron \cite{HoPi}
and later, Burakovsky \cite{BuHo} considered the extra time variable as some kind of universal parameter.  Additionally
in \cite{KoWi} the authors find a Schwarzschild-like solution to the Einstein equations for a metric of signature $-1$ and
make the identification between the extra time and electric charge.  In \cite{Wes} the second time is posited as the
inertial mass of a particle.  Furthermore, in \cite{DvGaSe}, the assumption
is made that the extra temporal dimension is a bulk dimension in which standard model fields are localized to one
particular instant in this extra time.

Regardless of the specifics of the extra-dimensional setup, it is natural to look for the extra-dimensional 
generalizations of the known four-dimensional solutions of the Einstein equations.  The black string solution of 
Horowitz and Strominger \cite{HoSt} is the ten-dimensional generalization of the Reissner-Nordstr\"{o}m  solution. 
The Kerr solution finds its five-dimensional counterpart in the Myers-Perry solution \cite{MyPe}.  In addition, 
higher-dimensional solutions have been found that have no counterpart in four-dimensional spacetime.  One example is the
five-dimensional black ring discovered by Emparan and Reall \cite{EmRe}.  The event horizon has a topology of $S^1 \times S^2$ 
and represents the first such solution found with nonspherical event horizon topology.  This hints at the fact that other solutions
similar to black holes may exist in higher dimensions.     

In this note we solve the five-dimensional vacuum Einstein equations, where no assumption is made as to the character of the 
extra dimension.  For simplicity, we look for solutions that are spherically symmetric, static and independent of the extra 
dimension.  We find such a solution and compare it with the four-dimensional Schwarzschild solution.  It is in this comparison
that we borrow from braneworld models the idea that test particles cannot access the bulk but instead are confined to the brane.
We will later refer to this as the brane condition.

The outline for the rest of the paper is as follows.  In Section II we present the equations and find solutions.  
Section III is given for comparison with the Schwarzschild solution.  We conclude with a discussion in Section IV.  
      
\section{Equations and Solutions}
\label{sec:equations}
We start by considering the following five-dimensional line element
\begin{equation}
ds^2 = Ae^{2a}dw^2 - e^{2b}dt^2 + e^{2c}dr^2 + r^2 d{\theta}^2 + r^2{\sin}^2{\theta}d{\phi}^2,
\end{equation}
where $a$, $b$, and $c$ are functions of $r$ only, $w$ is the extra dimension and $A$ (where $A^2 = 1$) is left
general in order to leave the character of the coordinate $w$ general.  
It is easily seen that this metric is spherically symmetric as well as static.
Using this metric assumption, the Einstein equations can be written as
\begin{equation}
A\left(-b'' -{b'}^2 + b'c' - {2(b' - c') \over r} + {e^{2c} - 1 \over r^2} + \Lambda e^{2c}\right) = 0
\end{equation}
\begin{equation}
-a'' -{a'}^2 + a'c' - {2(a' - c') \over r} + {e^{2c} - 1 \over r^2} + \Lambda e^{2c} = 0
\end{equation}
\begin{equation}
a'b' + {2(a' + b') \over r} - {e^{2c} - 1 \over r^2} - \Lambda e^{2c} = 0
\end{equation}
\begin{equation}
a'' + {a'}^2 + a'b' - (a' + b')c' + b'' + {b'}^2 + {a' + b' - c' \over r} - \Lambda e^{2c} = 0,
\end{equation}
where $'$ denotes differentiation with respect to $r$ and $\Lambda$ is the cosmological constant.  It can be shown
that $(5)$ is exactly satisfied if $(2)$, $(3)$, and $(4)$ hold, a result of the contracted Bianchi identities.  
Thus we have three independent equations for the three unknown functions $a$, $b$ and $c$.  It should also be noted that
the parameter $A$ only shows up as a multiplicative factor in equation $(2)$.  Thus the Einstein equations in vacuum are
completely independent of the character of the extra dimension.  This would obviously change if we demanded that the 
metric coefficients were dependent on $w$.  For this study, we simply divide out the $A$ in equation $(2)$ and the following
results are valid regardless of the character of $w$. 

Combining $(2)$ and $(3)$ and integrating by parts gives $a$ in terms of $b$ as
\begin{equation}
a = b + \lambda + \int{{\mu \over r^2}e^{(c-a-b)}dr},
\end{equation}
where $\lambda$ and $\mu$ are integration constants.  In order to find a solution, we set $\mu = 0$.  We have no physical
motivation for doing so and note that this may severely constrain the subsequent solution found.  The 
relationship $a = b + \lambda$ can then be substituted into the field equations.  When this is done $(3)$ becomes
equivalent to $(2)$ while $(4)$ and $(5)$ take the form
\begin{equation}
{b'}^2 + {4b' \over r} - {e^{2c} - 1 \over r^2} - \Lambda e^{2c} = 0
\end{equation}
\begin{equation}
2b'' + 3{b'}^2 - 2b'c' + {2b' - c' \over r} - \Lambda e^{2c} = 0,
\end{equation}
respectively.  

In order to make progress solving for $b$ and $c$, we set $\Lambda = 0$.  After doing so, we add one sixth of $(2)$
and $(7)$ together with one third of $(8)$ and obtain
\begin{equation}
{b'' \over 2} - {b'c' \over 2} + {b'}^2 + {b' \over r} = 0.
\end{equation}
If $b' = 0$, then $b$ is a constant and, from the equations, $c$ is also constant.  Thus we have a five-dimensional
generalization of flat, Minkowski spacetime.  In an effort to find a solution with nontrivial curvature, we assume
that $b \neq constant$ and solve $(9)$ for $c'$.  Upon integrating the resulting equation, we obtain a relationship
between $c$ and $b$ of the form
\begin{equation}
c = \ln{(r^2b')} + 2b + \kappa,
\end{equation}
where $\kappa$ is a constant.

We can now substitute this form for $c$ into $(7)$.  Upon making the transformations $r = e^x$ and
$b = {u-x \over 2}$, $(7)$ takes the form 
\begin{equation}
(1 - {\alpha}^2e^{2u}){\bar{u}}^2 + 2(3 + {\alpha}^2e^{2u})\bar{u} - (3 + {\alpha}^2e^{2u}) = 0,
\end{equation}
where $\alpha = e^{\kappa}$ and $\bar{}$ denotes differentiation with respect to $x$.  Solving this quadratic for 
$\bar{u}$, making the substitution $f = \sqrt{3 + {\alpha}^2e^{2u}}$ and integrating yields
\begin{equation}
x = {1 \over 2}\ln{(f^2 - 3)} \pm {1 \over \sqrt{3}}\ln{{f - \sqrt{3} \over f + \sqrt{3}}} + \ln{\delta},
\end{equation}
where $\delta$ is an integration constant.  We now transform back to our original variable $b$ and coordinate $r$ 
to get the relationship
\begin{equation}
\alpha \delta e^{2b} = {\left({\sqrt{1 + {{\alpha}^2r^2 \over 3}e^{4b}} + 1 \over \sqrt{1 + {{\alpha}^2r^2 \over 3}
e^{4b}} - 1}\right)}^{\pm{1 \over \sqrt{3}}}.
\end{equation}
While this equation cannot be expressed as $b(r)$, we can express $r(b)$ as
\begin{equation}
r = {\pm \sqrt{3} \over \alpha e^{2b}\sinh{(\sqrt{3}[b + {1 \over 2}\ln{(\alpha \delta)}])}}.
\end{equation}    
Having done so we can express $c(b)$ as
\begin{equation}
e^{2c} = {9 \over {(3 \cosh{(\sqrt{3}[b + {1 \over 2}\ln{(\alpha \delta)}])} + 2 \sqrt{3} \sinh{(\sqrt{3}[b + 
{1 \over 2} \ln{(\alpha \delta)}])})}^2}.
\end{equation} 

We can set the values of the constants $\alpha$ and $\delta$ by demanding asymptotic flatness as well as agreement
with the Newtonian weak field limit.  Upon inspection, asymptotic flatness requires that as $r \rightarrow \infty$,
$b \rightarrow 0$.  This amounts to letting $\delta = {\alpha}^{-1}$.  In order to agree with the weak field 
limit we set $\alpha = \mp M^{-1}$, where $M$ is the geometric mass of the four-dimensional slice of spacetime
accessible to material objects.  Since $0 \leq r < \infty$, 
and considering only positive $M$ scenarios, we choose the negative solution of $(14)$ and note that $b < 0$.  
Upon inspection we notice that the function $r(b)$ is double-valued and has a global minimum at 
$r=r_{min}\approx 4.58M$.  This corresponds to a minimum value for $b$ given by $b_{min} = -{1 \over \sqrt{3}}
{\tanh}^{-1}\left({\sqrt{3} \over 2}\right)\approx -0.76$.  Thus we only consider values of $b$ between $0$ and $b_{min}$, corresponding
to values of $r$ between infinity and $r_{min}$.  Therefore our solution, in these coordinates, does not cover the entire manifold.  The spatial section 
$0 \leq r < r_{min}$ remains uncovered.
\section{Comparison with Schwarzschild}
\label{sec:comparison}
It should be noted that the line element can now be written solely in terms of $b$ instead of $r$ as
\begin{equation}
ds^2 = A{\gamma}^2 e^{2b} dw^2 - e^{2b} dt^2 + {9 M^2 \over e^{4b} {\sinh}^4{\left(\sqrt{3}b\right)}}db^2
+ {3 M^2 \over e^{4b} {\sinh}^2{\left(\sqrt{3}b\right)}}\left(d{\theta}^2 + {\sin}^2{\theta}d{\phi}^2\right)
\end{equation}
where $\gamma = e^{\lambda}$.
This form for the line element is not very useful in trying to compare the solution with Schwarzschild. A 
more useful way of writing the line element is in a ``mixed" form given by
\begin{eqnarray}
ds^2 & = & A{\gamma}^2 e^{2b} dw^2 - e^{2b} dt^2 + {9 \over {(3 \cosh{(\sqrt{3}b)} + 
2\sqrt{3} \sinh{(\sqrt{3}b)})}^2}dr^2 \\ \nonumber
 & & + r^2 d{\theta}^2 + r^2 {\sin}^2{\theta}d{\phi}^2,
\end{eqnarray}
and supplement this with a relationship between $r$ and $b$ given by
\begin{equation}
r = {-\sqrt{3}M \over e^{2b} \sinh{(\sqrt{3}b)}}.
\end{equation}
In order to compare our solution with the Schwarzschild solution, it is advantageous to write the
latter solution in the same mixed form.  If this is done, the Schwarzschild solution becomes
\begin{eqnarray}
ds^2 & = & -\left(1 + {2 \over \sqrt{3}}e^{2b}\sinh{(\sqrt{3}b)}\right)d{\tilde{t}}^2 + {\left(1 + {2 \over \sqrt{3}}
e^{2b}\sinh{(\sqrt{3}b)}\right)}^{-1}dr^2 \\ \nonumber
& & + r^2 d{\theta}^2 + r^2 {\sin}^2{\theta} d{\phi}^2,
\end{eqnarray}
where $\tilde{t}$ may differ from $t$ by a scaling factor.  Because this factor makes no difference in our analysis,
we set it to one and no longer differentiate between $t$ and $\tilde{t}$.  Due to the brane condition, the
$g_{ww}$ term of the metric will not be needed in our comparison.  Therefore, $A$ and $\gamma$ can be left arbitrary.  

Upon inspection of the two metrics, we note two differences.  First of all, at $r=2M$, the Schwarzschild 
solution contains a coordinate singularity.  This value of $r$ corresponds to the event horizon as well as the
surface of infinite redshift.  In our metric, the event horizon is located at the end of our coordinate 
patch, namely $r=r_{min}$.  This is again only a coordinate singularity as can be verified by calculating the 
Kretschmann invariant.  This surface is also reachable in finite proper time.  Therefore, there is the hope that the
solution could be maximally extended inside the horizon as is done with the Schwarzschild solution in Kruskal coordinates.  
At present, however, no extension has been found.  Secondly, our event horizon does not represent a surface of infinite
redshift, nor does one appear to exist in our coordinate patch.  

In order to easily compare the geodesic equations for both metrics, we will write them in the same mixed form as 
above. Because of our symmetry assumptions, the $\theta$ and $\phi$ components of the two geodesic equations are 
identical.  The $t$ and $r$ components, however, are different.  For simplicity, we consider the interval equation 
in place of the $r$ equation, as is typically done \cite{d'Inverno}.
The $t$ equation is given, for the Schwarzschild solution and our solution respectively, as
\begin{equation}
\ddot{t} + {2 e^{4b} {\sinh}^2{(\sqrt{3}b)} \over M(2 \sqrt{3}e^{2b}\sinh{(\sqrt{3}b)} + 3)} \dot{t}\dot{r} = 0
\end{equation}
and
\begin{equation}
\ddot{t} + {2 e^{2b} {\sinh}^2{(\sqrt{3}b)} \over M(2 \sqrt{3}e^{2b}\sinh{(\sqrt{3}b)} + 3 \cosh{(\sqrt{3}b)})}
\dot{t}\dot{r} = 0.
\end{equation} 
The Schwarzschild interval is given by
\begin{equation}
-(1 + {2 \over \sqrt{3}}e^{2b}\sinh{(\sqrt{3}b)}){\dot{t}}^2 + {(1 + {2 \over \sqrt{3}}e^{2b}\sinh{(\sqrt{3}b)})}^{-1}
{\dot{r}}^2 + r^2 {\dot{\theta}}^2 + r^2{\sin}^2{\theta}{\dot{\phi}}^2 = K,
\end{equation}
while our interval takes the form
\begin{equation}
-e^{2b}{\dot{t}}^2 + {9 \over {(3 \cosh{(\sqrt{3}b)} + 2\sqrt{3} \sinh{(\sqrt{3}b)})}^2} {\dot{r}}^2 + 
r^2 {\dot{\theta}}^2 + r^2{\sin}^2{\theta}{\dot{\phi}}^2 = K,
\end{equation}
where $K=0$ for null geodesics and $K=-1$ for timelike geodesics.

The $t$ equation in both systems takes the same form apart from the coefficient of the $\dot{t}\dot{r}$ term.  
In order for our metric to possess the same geodesics as the Schwarzschild solution, the coefficients must
be equal.  For most values of $b$, this is not the case.  However, we can compare the values of the two theories
by choosing a value for $M$.  If we assume that we are looking for the geodesics in our
solar system then $M = 1500$ meters.  With this value for $M$ we find that outside the surface of the Sun the coefficients
agree to within one part in $10^{20}$, well beyond current experimental accuracy.  In this same instance, the 
${\dot{t}}^2$ and ${\dot{r}}^2$ coefficients in the interval equation agree to within one part in $10^{13}$ and four
parts in $10^{7}$, respectively.  Considering geodesics near the Earth, $M= 0.00445$ meters, the agreement, in the same 
order, is one part in $10^{26}$, less than one part in $10^{30}$ and two parts in $10^9$, respectively.   

\section{Discussion}
\label{sec:discussion}
We have assumed the existence of an extra dimension and solved the Einstein equations in this context.  We
have found a static, spherically symmetric solution to the equations that possesses the same geodesics as the 
Schwarzschild solution, within current experimental bounds.  Other features are out of line with the Schwarzschild
solution.  The event horizon for this solution exists at $r \approx 4.58M$ and in this coordinate system there does
not exist a surface of infinite redshift.  However, the experimental evidence for the size of the event horizon as 
well as the existence of an infinite redshift surface is at present nonexistent.  Therefore, it appears that the
physical realization of this solution cannot be immediately dismissed by current experimental evidence.  Thus, this
appears to be a valid alternative to the usual Schwarzschild solution.

We must temper this enthusiasm with some remarks on the shortcomings of this solution.  First, we have not found
a coordinate system that covers the entire spacetime.  Specifically, we cannot comment on the interior of the event
horizon nor on the presence of a curvature singularity.  Second, we have assumed a static solution from the outset.  
There is no a priori reason to believe that this solution is stable.  If it is not, it is very unlikely that this solution
would describe the final state of gravitational collapse.  We also must admit that this solution may depend crucially 
upon our choice of $\mu = 0$.  If this is not permissible we must see whether or not we obtain a solution with similar
properties keeping $\mu$ nonzero.  We also note that our assumption of $\Lambda = 0$ is not an assumption of braneworld models.  
Finally, we mention that demanding dependence of our metric coefficients on $w$ may
also destroy the solution.  More work must obviously be done to determine the significance of this solution.          

\section*{Acknowledgments}
\label{sec:ack}
I would like to thank Eric Hirschmann for useful discussions as well as for suggesting a change of variables that 
eventually led to the solution.


\begin{references}
\bibitem{Halpern} P. Halpern, {\em The Great Beyond} 
(John Wiley and Sons Inc 2004).

\bibitem{KaKl} T. Kaluza, {\em Sitzungober. Preuss. Akad. Wiss. Berlin}
p. 966 (1921); O. Klein, {\em Z. Phys.}  {\bf 37} 895 (1926).

\bibitem{HorWit} P. Horava and E. Witten, {\em Nucl. Phys. B}, {\bf 460}, 506 (1996).

\bibitem{ADD} N. Arkani-Hamed, S. Dimopoulos, G. Dvali, {\em Phys. Lett. B} {\bf 429}, 263 (1998).

\bibitem{RanSun} L. Randall and R. Sundrum, {\em Phys. Rev. Lett.} {\bf 83}, 4690 (1999);
                 L. Randall and R. Sundrum, {\em Phys. Rev. Lett.} {\bf 83}, 3370 (1999).

\bibitem{DvGaPo} G. Dvali, G. Gabadadze and M. Porrati, {\em Phys. Lett. B} {\bf 485}, 208 (2000).

\bibitem{KoWi} J. Kocinski and M. Wierzbicki, {\em Relativity, Gravitation, Cosmology}
(Nova Science 2004), gr-qc/0110075.

\bibitem{Bars}  I. Bars, {\em Class. Quant. Grav.} {\bf 18}, 3113 (2001).  

\bibitem{HoPi}  L. Horwitz and C. Piron, {\em Helv. Phys. Acta} {\bf 46}, 316 (1973).

\bibitem{BuHo}  L. Burakovsky and L. Horwitz, {\em Gen. Rel. Grav.} {\bf 27} 1043 (1995).

\bibitem{Wes} P. Wesson, {\em Phys. Lett. B} {\bf 538} 159 (2002).

\bibitem{DvGaSe} G. Dvali, G. Gabadadze and G. Senjanovic, hep-ph/9910207

\bibitem{HoSt} G. Horowitz and A. Strominger, {\em Nucl. Phys. B} {\bf 360} 197 (1991).

\bibitem{MyPe} R. Myers and M. Perry, {\em Ann. Phys. (N.Y.)} {\bf 172} 304 (1986).

\bibitem{EmRe} R. Emparan and H. Reall, {\em Phys. Rev. Lett.} {\bf 88}, 101101 (2002).

\bibitem{d'Inverno} R. d'Inverno, {\em Introducing Einstein's Relativity},
Oxford University Press (1992).

\end{references}
\end{document}